\documentclass{ifacconf}
\usepackage{array}
\usepackage{subcaption}
\usepackage{multirow}
\usepackage{inputenc} 
\usepackage{amsfonts} 
\usepackage{amssymb}
\usepackage{amsmath}
\usepackage{graphicx}      
\usepackage{natbib}        
\usepackage{algorithm}
\usepackage{algpseudocode} 
\makeatletter
\algrenewcommand\ALG@beginalgorithmic{\scriptsize}
\graphicspath{{fig/}}

\begin{document}
\begin{frontmatter}

\vspace{-0.3cm}
  \title{Toolbox for Discovering Dynamic System Relations via TAG Guided Genetic Programming}


\author[First]{Ștefan-Cristian Nechita} 
\author[First,SZTAKI]{Roland Tóth} 
\author[First]{Dhruv Khandelwal} 
\author[First]{Maarten Schoukens}


\address[First]{Department of Electrical Engineering
Eindhoven University of Technology, Eindhoven University of Technology, Eindhoven, The Netherlands (\{s.c.nechita, r.toth, d.khandelwal, m.schoukens\}@tue.nl).}
\address[SZTAKI]{Systems and Control Laboratory, Institute for Computer Science and Control, Kende u. 13-17, H-1111 Budapest, Hungary.}  

\begin{abstract}
Data-driven modeling of nonlinear dynamical systems often require an expert user to take critical decisions a priori to the identification procedure. Recently an automated strategy for data driven modeling of \textit{single-input single-output} (SISO) nonlinear dynamical systems based on \textit{Genetic Programming} (GP) and \textit{Tree Adjoining Grammars} (TAG) has been introduced. The current paper extends these latest findings by proposing a \textit{multi-input multi-output} (MIMO) TAG modeling framework for polynomial NARMAX models. Moreover we introduce a TAG identification toolbox in Matlab that provides implementation of the proposed methodology to solve multi-input multi-output identification problems under NARMAX noise assumption. The capabilities of the toolbox and the modelling methodology are demonstrated in the identification of two SISO and one MIMO nonlinear dynamical benchmark models.
\end{abstract}


\end{frontmatter}

\section{Introduction}
\vspace{-0.4cm}
Control design for complex dynamical systems rely heavily on accurate system models. A way to obtain such models is through first principle modeling. While this method provides generic models with clear physical interpretation it requires a considerable amount of time and user expertise. Another way to model the dynamical behaviour of a system is through data-driven system identification. Within this field there are numerous methods that require the user to take critical decisions (e.g. precisely selecting the model structure within prediction error methods (PEM)). In contrast, the system identification machine
learning strategies can automatically select or define model structures and features. The non-parametric machine learning methods such as Gaussian Process based Bayesian Estimators \cite{pillonetto_RKHS:2014}, Support vector machines (SVM) \cite{ming_guang_study_SVM:2004} and Artificial Neuron Networks (ANN), \cite{Goodfellow_et_al:2016}, \cite{billings_nonlinear:2013} describe large model spaces that can represent complex dynamical MIMO structures. However, often the obtained models via these methods lack interpretability and fail to provide generalization to unseen data or opacity regions of the system. On the other hand, the parametric machine learning methods, also known as symbolic regression, such as Tree Adjoining Grammar Guided Genetic Programming (TAG3P) \cite{Dhruv_thesis:2020}, and Equation discovery (EQ) \cite{Ferariu_patelli:2009} perform automated structure selection and yield time-domain solutions that directly represent the temporal modes of the system. In the doctoral thesis \cite{Dhruv_thesis:2020}, the author proposes a convenient way of defining the model set searching space through a novel Tree Adjoining Grammar modelling framework and conveys the critical decision of selecting the right model structure into a automated evolutive procedure based on Genetic Programming. Moreover, this thesis shows how the proposed method can discover physical relation directly from data (Duffing oscilator). This latest development with respect to the modeling framework focused on the single-input single-output (SISO) polynomial NARMAX model set but also included a considerable amount of variation (e.g. ability to embed $\mathrm{sin(\cdot)}$, $\mathrm{cos(\cdot)}$ or $\mathrm{abs(\cdot)}$ nonlinear operators and TAG representation of Box-Jenkins models).\\
The current paper work focuses on a novel grammar that extends the TAG modelling framework to multi-input multi-output (MIMO) polynomial NARMAX models. It is common for dynamic systems to have output channels with coupled dynamics. Our main contribution is defining a framework where the multi-output candidate models are represented by only one compact syntactic tree. By this, the dynamic modes are created, evolved and parametrized with respect to all output signals at once, thus considering the probable output dynamic coupling. Moreover, as our second contribution, an identification Matlab toolbox, which is publicly available on: \texttt{github.com/stefan-nechita/TAG\_Toolbox}, is provided. Using the toolbox, the user can easily select the structure searching space in terms of NARMAX (sub) model set(s) and also with custom made nonlinear building blocks. We have validated the modeling framework and Matlab implementation on two SISO and one MIMO nonlinear benchmark models. \\
The paper is structured as follows. Section \ref{section:Preliminaries} details the novel TAG modeling framework. Section \ref{section:Dual optimisation problem} describes the optimisation approach that drives the automated GP structure search procedure that is introduced in Section \ref{section:Genetic_programming_algorithm}. Section \ref{section:Results} shows the identification results of several benchmark models. In Section \ref{section:Conclusions} we draw conclusion on our results and present several future research direction. 
\vspace{-0.2cm}
\section{Model Structure via TAG}
\label{section:Preliminaries}
\vspace{-0.35cm}
The symbolic regression identification problem consists of determining an appropriate dynamic structure and corresponding parameters of a data generating system. The solution space is described as $\mathcal{S} = \mathcal{W} \times \mathcal{P}$, where $\mathcal{W}$ is the structure space and $\mathcal{P} \in \mathbb{R}^n $ is the parameter space, with $n$ arbitrary large, but finite. Hence naturally, a dual-optimization problem arises. For the proposed identification approach, TAG is used to describe the structure space $\mathcal{W}$. This chapter presents briefly the TAG modeling framework followed by a novel grammar proposal for MIMO polynomial NARMAX models. 
\begin{figure}[t]
\centering
\includegraphics[width=0.48\textwidth]{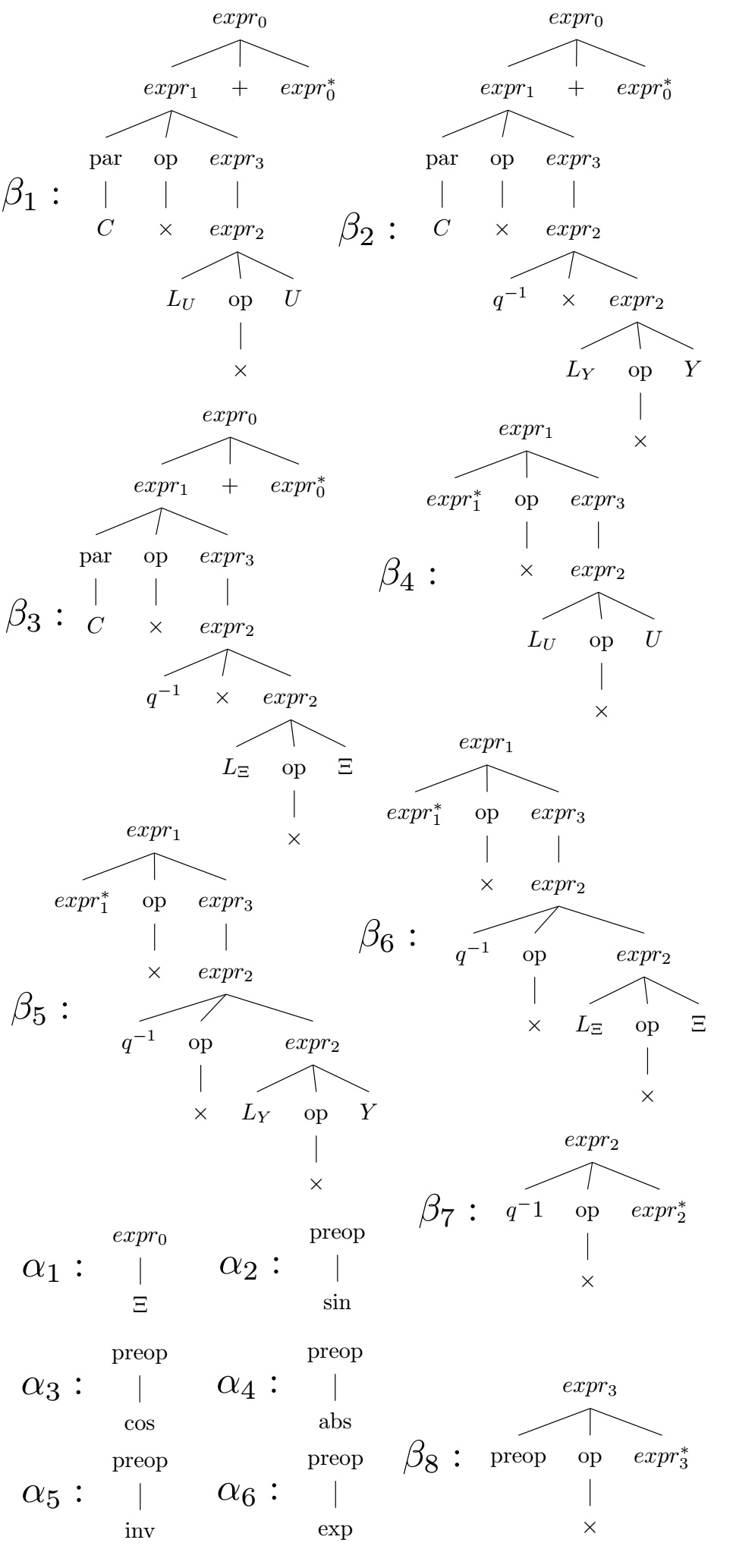}
\caption{Elementary trees $I \cup A$ of the extented $G_\mathrm{NARMAX}$}
\label{fig:Grammar_MIMO_e-NARMAX}
\end{figure}
\begin{table}[t]
\begin{center}
\captionsetup{width=9cm}
\caption{ Sub model sets included in $G_\mathrm{NARMAX}$}
\begin{tabular}{ccc}
Sub model & Grammar & Elementary trees \\ \hline
Input Poly. & $G_\mathrm{IP}$ & $\beta_1,\beta_4,\alpha_1$  \\
LTI & $G_\mathrm{LTI}$ & $\beta_1,\beta_2,\beta_7,\alpha_1$  \\
poly-NARX & $G_\mathrm{NARX}$ & $\beta_1,\beta_2,\beta_4,\beta_5,\beta_7,\alpha_1$  \\
ext-NARX & $G_\mathrm{extNARX}$ & $\beta_1,\beta_2,\beta_4,\beta_5,\beta_7,\beta_8,\alpha_{1,2,3,4}$\\
exp-NARX & $G_\mathrm{expNARX}$ & $\beta_1,\beta_2,\beta_4,\beta_5,\beta_7,\beta_8,\alpha_{1,5,6}$\\ \hline
\label{table:sub_model_Grammars}
\vspace{-0.2cm}
\end{tabular}
\end{center}
\end{table}
For a complete definition see \cite{kallmeyer:2009} and \cite{Dhruv_thesis:2020}.
In short, a candidate model described by a TAG can be seen as an orientated graph encoded by its derived tree $\gamma$, that has a root node $v_\mathrm{r}$, edges to its intermediate nodes $v_\mathrm{int}$ and leafs $v_\mathrm{l}$ all arranged in a purely (one to many) top to bottom fashion. The derived tree $\gamma$ is constructed based on its derivation tree $\Gamma_\gamma$. The later is formed by orientated (ordered) connections of elementary trees ($\beta$ and $\alpha$). The elementary trees are the "building blocks" of any TAG tree structure. In case of system identification, the correspond to elementary algebraic operations for signals, applying time operator such as: time shift (e.g. $\beta_7$) and elementary nonlinear functions such as $\beta_8$ with $\alpha_1 \hdots \alpha_6$. Alongside with label sets, the elementary tree form a TAG $G$. The structure of the elementary trees defines the rules that a certain grammar imposes over the shape of the derived trees $\gamma$ (i.e. it defines what is a model structure that can be generated from the elementary operations). Each such derived tree $\gamma$ represents a function $\mathcal{F}_\gamma$ via an interpreter function $\mathcal{E}(\gamma)$ that transposes the tree structure into the mathematical function $\mathcal{F}$. In our context, the design of the elementary trees defines the TAG language $\mathfrak{L}(G)$ (all the trees $\gamma$ that can be generated) thus, it directly defines the model set where $\mathcal{F}_\gamma = \mathcal{E}(\gamma)$ represents a model structure. Therefore elementary trees can be designed such that a TAG can represent, via it's language, an entire model set. TAG's are highly important as they allow to encode valid model representations and can seriously increase efficiency of GP based system identification.
\vspace{-0.35cm}
\subsection{TAG p-NARMAX modeling framework}
\vspace{-0.3cm}
Within this paper we focus on discrete-time MIMO polynomial NARMAX model set. Such a noise structure often provides enough flexibility to represent many dynamic systems in practice.  Further, we consider systems of form:
\begin{equation}
\begin{array}{ll}
Y(k) =& \mathcal{F}(\lbrace u_i(k-j) \rbrace_{j=1}^{n_u},\lbrace y_i(k-m) \rbrace_{m=1}^{n_y}, \\ & \lbrace \xi_i(k-l)\rbrace_{l=1}^{n_s}), i \in \mathrm{r}_{\lbrace u,y,\xi \rbrace}
\end{array}
\label{eq:def:MIMO_p_NARMAX_general_form}
\end{equation}
where
$U(k)$, $Y(k)$ and $\Xi(k)$ are input, output and process noise signals respectively with dimension $\mathrm{r}_{\lbrace u,y,\xi \rbrace} \times 1$, $\mathrm{r}_{\lbrace u,y,\xi \rbrace} \in \mathbb{N}$ and $n_u$, $n_y$ and $n_\xi$ are finite discrete time-delays with $n_u , n_\xi \in \mathbb{N}\cup \lbrace 0 \rbrace$, $n_y \in \mathbb{N}$ and $k \in \lbrace 1\hdots \mathrm{N} \rbrace$ finite number of time samples. If the case (\ref{eq:def:MIMO_p_NARMAX_general_form}) is restricted to polynomial relations, a suitable way to represent (\ref{eq:def:MIMO_p_NARMAX_general_form}) for TAG modeling framework, is as follows:
\vspace{-0.1cm}
\begin{equation}
\begin{array}{l}
\hspace{-0.1cm} Y(k)= \sum\limits^{p}_{i=1}C_i
    \prod\limits^{q_u}_{j=0}
    \prod\limits^{b_{i,j}}_{s_u}
    {L_{U,i,j}U(k-j)} \times \\ 
\hspace{-0.1cm}    
    \prod\limits^{q_y}_{m=1}
    \prod\limits^{a_{i,m}}_{s_y}
    {L_{Y,i,m}Y(k-m)}     \prod\limits^{q_\xi}_{l=1}
    \prod\limits^{d_{i,l}}_{s_\xi}
    {L_{\Xi,i,l}\Xi(k-l)} 
    +\Xi(k) \\

\end{array}
\label{eq:def:MIMO_p_NARMAX_TAG_suitable_form}
\end{equation}
\vspace{-0.cm}
where $L_{\lbrace U,Y,\Xi \rbrace}$ is a so called \textit{linking array} defined as:
\begin{equation}
\begin{array}{l}
L_{X} \in \mathbb{R}^{1 \times \mathrm{r}}, \mathrm{r} = \mathrm{dim}(X),  L = \begin{bmatrix} \mathrm{l}_{i} \end{bmatrix}_{i=1}^{\mathrm{r}}, \mathrm{l}_{i} \in \lbrace 0, 1 \rbrace\\
L_{X} \neq 0_{1 \times \mathrm{r}}
\end{array}
\label{eq:def:linking_array}
\vspace{-0.1cm}
\end{equation}
and $p \in \mathbb{N}$. The operation: $\prod\limits^{g_i}_{s=1} {L_{X,i,s}X(k-i)}$ is defined as a right hand side matrix multiplication with $\prod\limits^{0}_{s=1} {L_{X,i,s}X(k-i)} = 1$, where $X(k-i)$ is the value of signal $X$ at time moment $k-i$, $s$ is a selector operator counter, $L_{X,i,s}$ is a random linking array generated by (\ref{eq:def:linking_array}) and $g_i$ is the amount of right hand side multiplication of $X(k-i)$ with itself (e.g. right hand side matriceal rising to power: $X(k-i)^{g_j}$). 
The form (\ref{eq:def:MIMO_p_NARMAX_TAG_suitable_form}) can represent polynomial terms considering as variables all data channels and their time-shifted representatives $u_i(k-j)$, $y_i(k-m)$ and $\xi_i(k-l)$. 
Therefore, a given function $\mathcal{F(\cdot)}$ within the model set (\ref{eq:def:MIMO_p_NARMAX_TAG_suitable_form}) can be represented by a derived tree $\gamma$. 
\begin{prop}{TAG for MIMO p-NARMAX models\\}
Let $G_\mathrm{NARMAX}$ = $\langle N, T, S, I, A \rangle$ be a TAG with
\begin{itemize}
\vspace{-0.1cm}
\item $N = \lbrace expr_0, expr_1, expr_2, \mathrm{op}, \mathrm{par}\rbrace$,
\item $T = \lbrace U, Y, \Xi, +, C, \times, q^{-1}, L_Y, L_U,L_\Xi \rbrace$ , where $L_Y$, $L_U$ and $L_\Xi$ are "linking arrays", $U$, $Y$, $\Xi$ are the input, output and output noise signals and $C$ the is parameters vector.
\item $S = \lbrace expr_0 \rbrace$,
\item $I = \lbrace \alpha_1 \rbrace$,
\item $A = \lbrace \beta_1, \beta_2, \beta_3, \beta_4, \beta_5, \beta_6, \beta_7 \rbrace$, where the elementary trees $\beta_i$ and $\alpha_1$ are depicted in Figure \ref{fig:Grammar_MIMO_e-NARMAX}.
\end{itemize}
\vspace{-0.2cm}
The model set $M(G_{\mathrm{NARMAX}})$ represents the set of all polynomial models defined by Equation (\ref{eq:def:MIMO_p_NARMAX_TAG_suitable_form}) with $p, n_y, n_\xi \in \mathbb{N}$ and $n_u \in \mathbb{N}\cup\lbrace 0 \rbrace$. 
\label{proposal:MIMO_NARMAX_grammar}
\end{prop}
\vspace{-0.3cm}
Proposition \ref{proposal:MIMO_NARMAX_grammar} represents our main contribution over the TAG based modeling framework. As described in \cite{Dhruv_thesis:2020}, the TAG that represents the polynomial NARMAX model set can be enhanced or extended by considering $\mathrm{sin}(\cdot)$, $\mathrm{cos}(\cdot)$, $\mathrm{abs}(\cdot)$, $\mathrm{inv}(\cdot)$ and $\mathrm{exp}(\cdot)$ functions over the polynomial variables enlisted above. This modeling extension is enabled in the TAG modeling framework by considering the $\beta_8$ auxiliary tree and $\alpha_{2\hdots6}$ initial trees depicted in the lower part of Figure \ref{fig:Grammar_MIMO_e-NARMAX}. 
Similarly other functions can be added. Moreover, sub model sets included in $G_\mathrm{NARMAX}$ can be considered by selecting specific constituent elementary trees. Further extensions to the existed noise structure can be directly achieved as discussed in \cite{Dhruv_thesis:2020} by extending the elementary trees with further elements over the noise structre. A list of useful model sets is shown in Table \ref{table:sub_model_Grammars}. The user can select the constituent elementary trees (selecting a model set searching space) by selecting them in the toolbox file: \texttt{TAG\_MandatoryDefinition.m}. New elementary trees can be designed following the patterns in \texttt{CreateAuxTree.m} and \texttt{CreateInitTree.m}.
\vspace{-0.2cm}
\section{Identification Problem}
\vspace{-0.3cm}
\label{section:Dual optimisation problem}
Given a flexible model structure we would like to obtain an estimate of the underlying data generating system by finding a structure form with adequate complexity to achieve a desired level of approximation. This minimization can be formally defined as a dual optimization problem. 
Consider a TAG $G_{\mathrm{Model}}$ and it's equivalent model set $\mathcal{W}_{\mathrm{Model}}$ and a data generating system $\mathcal{F}_{\gamma_0}(\theta_0)$ described by a tree $\gamma_0 \in \mathfrak{L}(G_{\mathrm{Model}})$ with the real parameters $\theta_0$ that yield the real output sequence $Y_0(w_{\gamma_0} \vert \theta_0,D_\mathrm{N}) = Y_0(k)$, where $D_{\mathrm{N}} = \lbrace U(k), Y_{0}(k) \rbrace_{k=1}^\mathrm{N}$ is a data set of length $\mathrm{N}$ with $U(k)$ input sequence and $Y_{0}(k)$ stochastic response. Let $\mathcal{F}_{\hat\gamma}(\hat{\theta})=\mathcal{E}(\hat{\gamma})$ be a candidate model represented by $\hat{\gamma}$ tree and its assigned set of parameters $\hat{\theta}$. For the data set $D_{\mathrm{N}}$ the model $\mathcal{F}_{\hat{\gamma}}(\hat{\theta})$ yields the one step ahead prediction response $\hat{Y}_\mathrm{p}(w_{\hat{\gamma}} \vert \hat{\theta},D_\mathrm{N}) = \hat{Y}_\mathrm{p}(k)$ and simulation response $\hat{Y}_\mathrm{s}(w_{\hat{\gamma}} \vert \hat{\theta},D_\mathrm{s,N}) = \hat{Y}_\mathrm{s}(k)$, where $D_\mathrm{s,N} = \lbrace U(k), \hat{Y}_{\mathrm{s}}(k) \rbrace_{k=1}^\mathrm{N}$. The two responses generate an error point $E=(E_{\mathrm{s}},E_{\mathrm{p}})\in \mathbb{R}^2$ where $E_{\mathrm{s}}$ is the root mean square simulation error ($\mathrm{RMS}_\mathrm{s}$) produced by $\hat{Y}_{\mathrm{s}}(k)$ and $E_{\mathrm{p}}$ is the root mean square prediction error ($\mathrm{RMS}_\mathrm{p}$) produced by $Y_{\mathrm{p}}(k)$. The main aim of the identification strategy is to minimize the error point $E$. Therefore the identification procedure searches for the solution of the following dual optimization problem:
\begin{equation} 
\begin{array}{lll}
\underset{w_{\gamma}}{\text{min}}	
&
J(w_{\gamma},\hat{\theta})= 
&
 \mathrm{min} \left( E \left(w_\gamma, \hat{\theta} \right)  		
			   \right) \\
 \text{s.t.} 						
 &					    
 &									    
 \\
 \hat{\theta} =  \underset{\theta}{\text{min}} 
 & 
 J_{\mathrm{sub}}(\theta)= 
 &
\omega_{\mathrm{s}}E_{\mathrm{s,\tau}} ( \theta )+  \omega_{\mathrm{p}}E_{\mathrm{p}} ( \theta )			
\end{array}
\label{eq:minimisation_problem}
\end{equation}
\begin{equation}
\begin{array}{l}
\hspace{-0.25cm}
E_{\mathrm{s}}(\theta) = \frac{1}{\mathrm{r}_y}\sum\limits_{\mathrm{i}=1}^{\mathrm{r}_y}\sqrt{\frac{1}{\mathrm{N}}\mathrm{e}^{\top}_{\mathrm{i,s}} \mathrm{e_{i,s}}}, \quad \hspace{-0.2cm} E_{p}(\theta)=	\frac{1}{\mathrm{r}_y}\sum\limits_{\mathrm{i}=1}^{\mathrm{r}_y} \sqrt{\frac{1}{N} \mathrm{e}^{\top}_{\mathrm{i,p}} \mathrm{e_{i,p}}}
\end{array}
\label{eq:Error_RMS_sim_pred}
\end{equation}
where
\begin{equation}
\begin{array}{lll}
\mathrm{e_{{i,\lbrace s,p \rbrace}}} & = & [ y_{0,\mathrm{i}}(k) - \hat{y}_{\mathrm{i,\lbrace s,p \rbrace}}(w_\gamma,k  \vert \hat{\theta},D_\mathrm{N}) ]_{k=1}^{\mathrm{N}},
\end{array}
\end{equation}
$\omega_{\mathrm{s}}$ is the simulation error weight and $\omega_{\mathrm{p}}$ is the prediction error weight. The weight values play a role in determining what parameter estimation procedure can be deployed to solve the sub-optimization problem. They will be further detailed later. The $\mathrm{RMS}_\mathrm{p}$ is the error produced by a candidate model that has access to the past real system input and output data ($U(k)$ and $Y_0(k-n_y)$) while the $\mathrm{RMS}_\mathrm{s}$ is the error produced by a candidate model that uses the real input signal $U(k)$ and past own simulated values of output signal $\hat{Y}_\mathrm{s}(k-n_y)$. Minimizing the $\mathrm{RMS}_\mathrm{p}$ enforces the candidate model to approximate the dynamical, self-feeding modes of the data generating system. In short, the $\mathrm{RMS}_\mathrm{s}$ is the metric that measures how well the candidate models performs autonomously and offers a much stronger indication with respect to how well the candidate model approximates the data generating system.

\vspace{-0.3cm}
\section{Estimation via Genetic programming}
\vspace{-0.3cm}
\label{section:Genetic_programming_algorithm}
 To solve the multi-objective dual optimization problem described above, we designed a Genetic Programming (GP) algorithm that evolves a population of tree structures through TAG designed crossover and mutation genetic operators, perform parameter estimation for each structure and sorts each generation based on two fitness criterion $\mathrm{RMS}_\mathrm{s}$ and $\mathrm{RMS}_\mathrm{p}$ using the multi-objective non-dominating sorting algorithm.
 \vspace{-0.3cm}
\subsection{Main Algorithm}
\vspace{-0.35cm}
 The main steps of the GP algorithm are presented in Algorithm \ref{alg:TAG3P}. The GP is initialized by defining the genetic parameters: population size ($\mathrm{Pop}$), number of generations ($\mathrm{Gen}$), number of maximum auxiliary trees that can be used in each derivation tree ($\mathrm{Complexity}$) and crossover parameter ($\mu \in [0-100\%]$). Inside the iterative loop, the crossover, mutation, interpreter function, parameter estimation, evaluation and non-dominating sorting procedures are executed sequentially in order to propose, construct, evaluate and sort new dynamical structures. At the end the solution is considered to be the first Pareto front of the last generation. Since within the Pareto solution the models do not dominate each other, in terms of the two considered fitness criterion, any of them can be selected as a final candidate model that minimizes problem (\ref{eq:minimisation_problem}). The Algorithm \ref{alg:TAG3P} can be found in \texttt{TAG3P\_main.m} file. Next we will explain the main procedures in detail. \\
\begin{algorithm}[t]
\caption{TAG GP main}
\label{alg:TAG3P}
\begin{algorithmic}
\State Define $\mathrm{Pop}$ \Comment{Define Population Size}
\State Define $\mathrm{Complexity}$ \Comment{Define maximum complexity}
\State Define $\mathrm{Gen}$ \Comment {Define the maximum number of generations}
\State $\mathrm{G(1)}$ $\gets$ RandomPopulation \Comment {Generate a random population of trees}
\State $\mathrm{G(1)}$ $\gets$ Interpreter($\mathrm{G(1)}$) \Comment{Construct the candidate model}
\State $\mathrm{G(1)}$ $\gets$ ParameterEstimation($\mathrm{G(1)}$) 
\State $\mathrm{G(1)}$ $\gets$ Evaluate($\mathrm{G(1)}$) \Comment{Compute $E_{\mathrm{s}}$ and $E_{\mathrm{p}}$ for $\mathrm{G(1)}$}
\While{$i\leq$ $\mathrm{Gen}$ }
	\State $\mathrm{Q_1}$ $\gets$ CrossoverOffsprings($\mathrm{G(i)}$) \Comment{$\mathrm{Card}( \mathrm{Q_1}) = \mathrm{Pop}$ }
	\State $\mathrm{Q_2}$ $\gets$ MutationOffsprings($\mathrm{G(i)}$)  \Comment{$\mathrm{Card}( \mathrm{Q_1}) = \mathrm{Pop}$ }
	\State $\mathrm{Q_{1,2}}$ $\gets$ Interpreter($\mathrm{Q_{1,2}}$) \Comment{see \texttt{CreateTreeFunction.m}}
	\State $\mathrm{Q_{1,2}}$ $\gets$ ParameterEstimation($\mathrm{Q_{1,2}}$) 
	\State  $\mathrm{Q_{1,2}}$ $\gets$ Evaluate( $\mathrm{Q_{1,2}}$) \Comment{Compute $E_{\mathrm{s}}$ and $E_{\mathrm{p}}$ for  $\mathrm{Q_{1,2}}$}
	\State $\mathrm{R}$ $\gets$ $\mathrm{G(i)}$ $\cup$ $\mathrm{Q_{1}}$ $\cup$ $\mathrm{Q_{2}}$
	\State $\mathrm{R}$ $\gets$ NSGA-II($\mathrm{R}$) \Comment{Sorting R into Pareto fronts}
	\State $\mathrm{G(i+1)}$ $\gets$ $\mathrm{R}(1:\mathrm{Pop})$ \Comment{Select the first $\mathrm{Pop}$ candidates from the \begin{flushright} first Pareto fronts of $\mathrm{R}$ \end{flushright}}
\EndWhile
\State \textbf{Save} $\mathrm{G}(\mathrm{Gen})$\Comment{collect the Pareto solution}
\end{algorithmic}
\end{algorithm}
\vspace{-0.6cm}
\subsection{Crossover and Mutation genetic operators}
\vspace{-0.3cm}
In Crossover, two parents (individuals of population) have their genotype combined in order to form new individuals called offsprings. Through crossover, no new information is added to the population. By switching strings of genotype between individuals, over generations, the genes that yield smaller fitness values tend to become more frequent in the population. In this way, a local exploration of the search space is performed. Consequently, via crossover, a population is exploring a local minimal point. As described in \cite{Hoai_TAG_language_GP:2003}, within TAG3P+ a sub-tree crossover is defined as follows. Two trees $\gamma_1$, randomly selected from the first $\mu$ structures of $\mathrm{G}(i)$, and $\gamma_2$, randomly selected from the entire $\mathrm{G}(i)$. A randomly chosen point in each of the two derivation trees is chosen, subject to the constraint that each sub-tree can be adjoined to the other parent tree. For each parent, the derivative tree $\Gamma$ is split in two $\Gamma_{\mathrm{STEM}}$ and $\Gamma_{\mathrm{TAIL}}$. The offsprings are created by adjoining the stem of the first parent with the tail of the second and vice versa. The \texttt{TAGCrossover.m} file hosts the implementation of the Crossover operator.\\
In Mutation, an offspring is proposed by eliminating or adjoining elementary trees starting from a derivation tree $\Gamma \in \mathrm{G}(i)$. In our implementation, for each structure of $\mathrm{G}(i)$ an offspring is created by mutation. By random addition or deletion of elementary trees to or from the parent derivation tree, the mutation operator is the procedure through which the evolution process performs global exploration of the searching space. The mutation genetic operator is implemented in \texttt{TAGMutation.m}. Both crossover and mutation functions are called inside the main loop in \texttt{TAG\_GP\_Step1.m}.
\vspace{-0.35cm}
\subsection{Parameter estimation procedures}
\vspace{-0.3cm}
Every model constructed through crossover, mutation and random generation requires optimization of its parameters to assess its accuracy in terms of (\ref{eq:minimisation_problem}). The parameter estimation can be performed with respect to both simulation and prediction error (non zero $\omega_\mathrm{s}$ and $\omega_\mathrm{p}$ weights e.g. swarm-optimization approach CMA-ES by \cite{hansen_CMA:2001}, see also \texttt{CMAES.m} file) or only prediction error ($\omega_\mathrm{s}=0$ and $\omega_\mathrm{p}=1$ (e.g. least square procedure see \texttt{ParEst\_LS.m} file). Considering both $\mathrm{RMS}_\mathrm{s}$ and $\mathrm{RMS}_\mathrm{p}$ in parameter estimation transforms the sub optimization problem into a non-convex optimization problem, making it difficult and time-consuming to solve. If only the prediction error is considered, any model defined by a function $\mathcal{F}_\gamma$ with $\gamma \in \mathfrak{L}(G_\mathrm{NARMAX})$ can be rewritten as (\ref{eq:LeastSquaresForm})
\begin{equation}
\begin{array}{l}
\Psi=\Phi\Theta + E_\Theta\\
\end{array}
\label{eq:LeastSquaresForm}
\end{equation}
where, for $p$ polynomial terms as described in (\ref{eq:def:MIMO_p_NARMAX_TAG_suitable_form}), $\hat{\Psi} \in \mathbb{R}^{\mathrm{N} \times n_y}$ is the model output data set, $\Phi \in \mathbb{R}^{\mathrm{N}\times p}$ is the evolution of each polynomial term over $D_\mathrm{N}$ and $\Theta \in \mathbb{R}^{p\times n_y}$ is the matrix corresponding to the parameter vector $\Theta$. The set of parameters that minimize the sub optimization problem (\ref{eq:minimisation_problem}) is computed as on  (\ref{eq:LeastSquaresParameters}).
\begin{equation}
\begin{array}{l}
\hat{\Theta} = \left( \Phi^\top\Phi \right)^{-1}\Phi^\top \Psi.
\end{array}
\label{eq:LeastSquaresParameters}
\end{equation}
Within this report, we have opted to used the least squares method for parameter estimation during the genetic evolution process because it is considerable faster than using parameter estimation methods that consider both simulation and prediction error. In the toolbox, the parameter estimation procedure is called inside the main loop in \texttt{TAG\_GP\_Step2.m}. Moreover, the toolbox user has the option to chose between three parameter estimation procedures: least squares, swarm optimization CMA-ES or unconstrained iterative method (see \texttt{ParEst\_fminunc.m}).
\vspace{-0.3cm}
\subsection{Multi-objective non-dominated sorting} 
\vspace{-0.3cm}
The evolution of dynamical structure as presented above can be guided by a multi-objective criterion. In the presented algorithm, we have considered only simulation and prediction error ($E_\mathrm{s}, E_\mathrm{s}$), but criterion like derivation tree complexity (see \cite{Dhruv_thesis:2020}) or cardinality of the set of parameters can also be included. In the toolbox, $E_\mathrm{s}$ and $E_\mathrm{s}$ values are computed in \texttt{TAG\_GP\_Step3.m} file. In the multi-objective genetic programming literature, most of the evolutionary strageties bases their findings on Pareto optimality criterion. We further present the Pareto dominance definition \cite{emmerich_tutorial:2018}.
\begin{defn}{\textit{Pareto dominance}}\\
Given two vectors in the objective space, $O^{(1)},O^{(2)} \in \mathbb{R}^{m}$, then the point $O^{(1)}$ said to \textit{Pareto dominate} the point $O^{(2)}$ ($O^{(1)} \prec_{Pareto} O^{(2)}$), if and only if $\forall i \in \lbrace 1,\ldots,m \rbrace : O^{(1)}_i \leq O^{(2)}_i$ and $\exists j \in \lbrace 1,\ldots,m \rbrace: O^{(1)}_j < O^{(2)}_j$. In case that $O^{(1)} \prec_{\mathrm{Pareto}} O^{(2)}$ the first vector is not worse in each of the objectives and better in at least one objective than the second vector.
\label{def:Pareto_dominance}
\end{defn}
\vspace{-0.2cm}
Based on the Pareto dominance $\prec_{\mathrm{Pareto}}$, one can group a set of candidates into fronts. Each candidate has a dominance level and it is based on the number of how many other candidates are Pareto dominated by it. A Pareto front, $F_i$, can be seen as a contour on which all the candidates have the same dominance level. The order of dominance sorts the Pareto fronts between themselves. The Pareto optimal solution is the front that has the highest dominance level, as known as the set of non-dominated solution. A way to construct the Pareto fronts for a given set of dynamical structures is the NSGA-II algorithm detailed in \cite{deb_fast:2002} (see \texttt{NSGAII.m}). The NSGA-II algorithm is called in \texttt{TAG\_GP\_Step4} file.
Every generation, for structure sorting procedure, the new models constructed through crossover and mutation are benchmarked against a distinct data set $D_{\mathrm{N}}^{\mathrm{test}}$.
\vspace{-0.3cm}
\section{Results}
\vspace{-0.3cm}
\label{section:Results}
We tested the TAG3P identification algorithm against two SISO and one MIMO benchmark models. For each model we considered three distinct data sets cathegories: $D_\mathrm{N}^\mathrm{est}$ for parameter estimation, $D_{\mathrm{N}}^\mathrm{test}$ for multi-objective sorting and $D_{\mathrm{N}}^\mathrm{val}$ for computing validation $\mathrm{RMS}_\mathrm{s}$ and $\mathrm{RMS}_\mathrm{p}$ metrics described in Equations (\ref{eq:Error_RMS_sim_pred}). These metrics are used to compare the results obtained through the proposed method with the ones presented in literature. For all benchmark systems, the comparison is shown in Table \ref{table:Results_and_comparison}. For each benchmark model, out of the Pareto solution, we have selected the candidate model that yields the lowest average simulation error over the $D_{\mathrm{N}}^\mathrm{test}$ data sets. For the MIMO benchmark model described in \cite{StirTank_TOTH:2010}, the authors measured their identification method performance in Best Fit Rate ($\mathrm{BFR}$). Thus, for the MIMO case, alongside the $\mathrm{RMS_s}$ value we have also computed a $\mathrm{BFR}$ metric.
\begin{table}[t]
\begin{center}
\captionsetup{width=9cm}
\caption{$\mathrm{RMS_s}$ and $\mathrm{RMS_p}$ results of TAG3P Matlab Toolbox over the benchmark models in comparison with other system identification strategies from the literature.}
\label{table:Results_and_comparison}
\vspace{-0.2cm}
\begin{tabular}{lcc}
\multicolumn{1}{c}{Bouc-Wen hysteresis model}&                    &                  \\ \hline
\textbf{TAG3P -} $G_\mathrm{NARX}$			& $6.52\mathrm{e-}5$ & $7.37\mathrm{e-}6$\\
Full PLNSS	\cite{BoucWen_ESFAHANI:2017}		& $1.20\mathrm{e-}5$ & -                 \\
Decoupled PLNSS	\cite{BoucWen_ESFAHANI:2017}	& $1.40\mathrm{e-}5$ & -                 \\ 
LMN - NARX \cite{BoucWen_BELZ:2017}			& -                  & $9.86\mathrm{e-}6$\\
LMN - NFIR	\cite{BoucWen_BELZ:2017}		& $1.63\mathrm{e-}4$ & -                 \\ \hline
\multicolumn{1}{c}{Coupled electric drive}	&                    &                   \\ \hline
\textbf{TAG3P -} $G_\mathrm{extNARX}$		& $1.28\mathrm{e-}2$ & $3.27\mathrm{e-}3$\\
TAG3P  \cite{Dhruv_thesis:2020}				& $1.2\mathrm{e-}1$  & $3.73\mathrm{e-}3$\\
GA + DE \cite{CEDrive_Ayala:2014}			& $1.8\mathrm{e-}1$  & $4.0\mathrm{e-}2$ \\ \hline
\multicolumn{1}{c}{Continuous Stirring Tank Reactor}					& $\mathrm{RMS_s}$	 & $\mathrm{BFR}$ 	 \\ \hline
\textbf{TAG3P -} $G_\mathrm{expNARX}$		& $1.6749$ 			 & $92.80\%$		 \\
LPV-OBF	\cite{StirTank_TOTH:2010}			& -                  & $97.54\%$	 		 \\ \hline
\end{tabular}
\end{center}
\end{table}
\begin{table}[ht]
\begin{center}
\captionsetup{width=9cm}
\caption{ TAG and genetic parameters used for the benchmark problem.}
\vspace{-0.2cm}
\begin{tabular}{ccccc}
Benchmark model	& TAG 				& $\mathrm{Pop}$&$\mathrm{Gen}$	&$\mathrm{Complexity}$	\\ \hline
BoucWen Osc.	& $G_\mathrm{NARX}$ & $36$ 			&$350$			&$150$					\\
Coupled El. drive	& $G_\mathrm{extNARX}$ & $50$ 			&$400$			&$150$				\\
Styr Tank model	& $G_\mathrm{expNARX}$ & $60$ 			&$350$			&$120$				\\ \hline
\label{table:Benchmark_results_set_ip}
\vspace{-0.2cm}
\end{tabular}
\end{center}
\end{table}
\vspace{-0.7cm}
\subsection{SISO benchmark models}
\vspace{-0.3cm}
\subsubsection{Bouc-Wen model}
The Bouc-Wen model has been used during the last decades to represent hysteretic effects in mechanical engineering. The current benchmark represents a Bouc-Wen model with synthetic input and output data. This system is challenging to identify for a series of reasons. On on which is that the system  possesses a dynamic nonlinearity that is governed by a non measurable internal variable. To identify this model we used the TAG and genetic programming parameters described in the second entry of Table \ref{table:Benchmark_results_set_ip}.\\
For the parameter estimation and testing data sets ($D_\mathrm{
N}^\mathrm{N}$, $D_{\mathrm{N}}^\mathrm{test}$) we have generated $5$ data sets of $\mathrm{N}=4096$ samples each using the algorithm indicated in \cite{BoucWen:2020}. The validation data set $D_\mathrm{N}^\mathrm{val}$ was considered the sine sweep data set provided by the authors.
\vspace{-0.3cm}
\subsubsection{Coupled electric drive model}
\begin{figure}
\centering
\vspace{-0.3cm}
\includegraphics[width=0.5\textwidth]{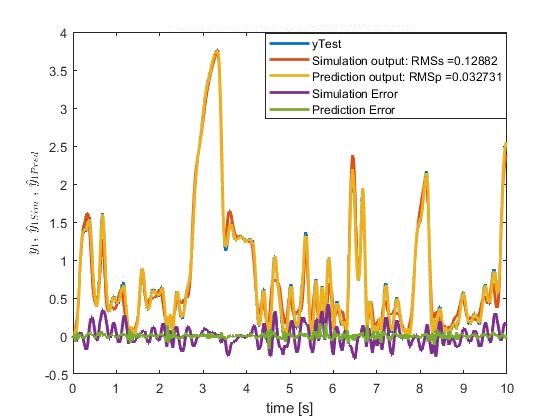}
\caption{Coupled electric drive system validation output compared with the simulation and prediction responses of the candidate model, simulation and prediction error on $D_{\mathrm{N}}^\mathrm{val}$}
\label{figure:results:CEdrive_model}
\end{figure}

The coupled electric drives consists of two electric motors that drive a pulley using a flexible belt. The pulley is held by a spring, resulting in a lightly damped dynamic mode.  The drive control for the pulley is designed only for tracking the speed reference signal. A pulse counter is used to measure the angular speed of the pulley. Thus, the sign of the velocity is unknown. The available data sets are short ($\mathrm{N}=500$), and together with the absolute value component of the velocity profile make this system interesting from an identification point of view. Because of the known absolute value behaviour of the output signal, to identify this model we used the extended TAG and genetic programming parameters enlisted in the third entry of Table \ref{table:Benchmark_results_set_ip}.\\
As described in \cite{CEDrive:2017}, the estimation data set $D^{\mathrm{est}}_\mathrm{N}$ contained the data expressed by $\mathrm{u11}$ as input and $\mathrm{z11}$ as output. The testing  $D_{\mathrm{N}}^\mathrm{test}$ and validation $D_{\mathrm{N}}^\mathrm{val}$ data sets contained the data expressed by $\mathrm{u12}$ as input and $\mathrm{z12}$ as output.\\
The identification results, for the same validation data set, are fairly similar to the TAG3P implementation (in Mathematica) described in \citep{Dhruv_thesis:2020}.
Figure \ref{figure:results:CEdrive_model} shows the sine sweep validation output signal, the simulated and predicted candidate model output and simulation and prediction error signals. 
\vspace{-0.35cm}
\subsection{MIMO benchmark model}
\vspace{-0.4cm}
\subsubsection{Continuous Stirred Tank Reactor model}
\begin{figure}[t]
\centering
\vspace{-0.2cm}
\includegraphics[width=0.5\textwidth]{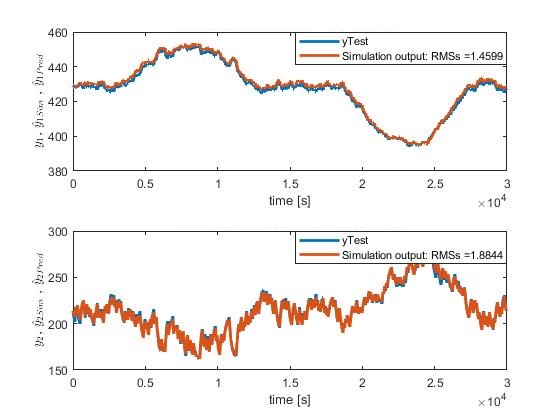}
\vspace{-0.6cm}
\caption{CSTR validation output compared with the simulation responses of the candidate model on $D_{\mathrm{N}}^\mathrm{val}$}
\label{figure:results:BoucWen_model}
\end{figure}
The main contribution of this paper is the extension of the TAG modeling framework to MIMO complex models. For this we tested the TAG3P MIMO identification procedure on an ideal, simulated, Continuous Stirred Tank Reactor (CSTR) that is fully described in \cite{StirTank_TOTH:2010}. In short, the CSTR resembles a chemical conversion of an inflow substance into a product. The chemical conversion is described by a highly nonlinear dynamic relation between input signals $U = [ \mathrm{Q}_1, \mathrm{T_c}, \mathrm{C_1}]^\top$ (input flow, coolant temperature and concentration of the inflow) and output signals $Y = [\mathrm{T_2}, \mathrm{C_2}]^\top$ (temperature in the reactor and concentration in the reactor). Since the benchmark model is fully known, $10$ estimation and testing data sets were generated. These have a length of $\mathrm{N}=1000$ and considered the signals $\mathrm{Q}_1$, $\mathrm{T_c}$ as a pseudo random binary signal (PRBS) form with values of $\pm 10\%$ of nominal values and $\mathrm{C_1}$ as a slow variation, starting from nominal, toward $10$ equidistant operational points in the interval $50 \% - 150 \%$ of nominal. In this way we have excited the system components around the operation values. Over the synthetic output signals $\mathrm{T_2}$ and $\mathrm{C_2}$, a uniformly distributed noise of amplitude $0.5$ and $2$ respectively have been added. This addition mimics a sensor signal to noise ratio of $63.68$ for $\mathrm{T_2}$ and $45.56$ $\mathrm{C_2}$. This experiment design did not consider potential costs of the materials if the input signals were to be applied to a real reactor. The data set $D_\mathrm{N}^\mathrm{val}$ ($\mathrm{N}=500$) was designed to test how well the candidate model describes the global behavior of the reactor and it is similar to the global validation data set presented in \cite{StirTank_TOTH:2010}. Because of the known inverse and exponential terms within the model equations, to identify this model we used the extended TAG and genetic programming parameters enlisted in the forth entry of Table \ref{table:Benchmark_results_set_ip}.
\vspace{-0.2cm}
\section{Conclusion and Future Work}
\vspace{-0.3cm}
\label{section:Conclusions}
As presented in the Table \ref{table:Results_and_comparison} the new Matlab implementation of the TAG3P identification strategy could identify the three SISO benchmark models with a various degrees of fidelity. The results for Bouc-Wen oscillator and Coupled Electric Drive show $\mathrm{RMS_s}$ and $\mathrm{RMS_p}$ values on par with other literature solutions while for the Parallel Wiener-Hammerstein model, the obtained $\mathrm{RMS_s}$ is smaller than the Best Linear Approximation but considerable larger than the specialized Parallel W-H solution proposed in \cite{PWH_SCHOUKENS:2015} by a factor of 3. In case of the MIMO CSTR system, the BFR metric shows that the proposed TAG modeling framework can obtain a valid model from data. In all cases, te TAG guided genetic programming can provide reliable candidates that represent complex SISO or MIMO nonlinear system. Moreover, it shows a fine trade-off between performance of the identified model and the amount of critical decision the user has to take. Nevertheless the paper introduced and made available the first version of the Matlab Toolbox for TAG3P identification strategy.\\
As future work, in terms of the modeling framework and model space selection, the current TAG MIMO framework and Matlab implementation offer enough flexibility for proposing a genetic programming guided identification procedure for systems that can be described by polynomial nonlinear state space models. The aim of such framework is to enable the genetic evolution to automatically select the dynamic structure and the number of states. 

\vspace{-0.3cm}
\bibliography{mybib}             
 \end{document}